\documentclass[usenatbib,useAMS]{mn2e}
\usepackage{times}
\usepackage{epsfig}

\begin{document}
\title{Can microlensing fold caustics reveal a second stellar limb-darkening coefficient?}
\author[M. Dominik]{M. Dominik \\
University of St Andrews, School of Physics \& Astronomy, North Haugh,
St Andrews, KY16 9SS}

\maketitle

\begin{abstract}
Dense high-precision photometry of microlensed stars during a fold-caustic passage can be
used to reveal their brightness profiles from which the temperature of the stellar 
atmosphere as function of fractional radius can be derived. While the capabilities of 
current microlensing follow-up campaigns such as PLANET allowed for several 
precise measurements of
linear limb-darkening coefficients, all attempts to reveal a second limb-darkening coefficient
from such events have failed. It is shown that 
the residual signal of a second coefficient characterizing
square-root limb darkening is $\sim\,$25 times smaller which prevents a proper determination
except for unlikely cases of very high caustic-peak-to-outside magnification
ratios
with no adequate event being observed so far or for source stars passing over a cusp
singularity. Although the presence of limb darkening can be well established from the data,
a reliable measurement of the index of an underlying power-law cannot be obtained.
\end{abstract}


\section{Introduction}

Since \citet{SW1987} have proposed to use 
caustics of gravitational lenses for measuring the brightness profile
of closely-aligned background sources, dense high-precision 
photometry on microlensing events involving fold-caustic passages
has provided measurements of
linear limb-darkening coefficients for several stars 
\citep{PLANET:M41, PLANET:O23,
joint, PLANET:EB5}.

For stellar brightness profiles that involve linear, square, or square-root terms in
$\cos \vartheta$, where $\vartheta$ is the emergent angle, 
properties of light curves during fold-caustic passages have been
studied by \citet{Rhie:LD}, while 
prospects and strategies for determining the linear limb-darkening 
coefficient have recently been discussed in detail by \citet{Do:FoldLD},
showing the ability of obtaining precise measurements even with moderate
use of the current capabilities.

This letter will however show that the measurement of additional coefficients, e.g.\
corresponding to a square-root law term, is not possible for typical events that involve
a fold-caustic passage. Attempts of such measurements \citep{PLANET:EB5}
have resulted in the meaningful measurement of a single
linear combination of the involved limb-darkening coefficients only, while
a similar result has been obtained for an event where the source passes over a cusp
\citep{Abe}, where however the exhibited differential magnification is modest
due to the source size being relatively large compared to the cusp.
In contrast, the cusp-passage event discussed by \citet{PLANET:M28} provides a better size ratio
between source and cusp and to date presents
the only measurement of both linear and square-root limb-darkening coefficients
by microlensing.

\section{Intensity profile and lightcurve}
\label{sec:lightcurve}

With $\overline{I}$ denoting the average brightness 
and $\rho$ the fractional radius of an observed source star, its 
brightness profile for a given filter reads 
\begin{equation}
I(\rho) = \overline{I}\,
\xi(\rho)\,,
\end{equation} 
where
$\xi(\rho)$ is a dimensionless function fulfilling the
normalization
\begin{equation}
\int\limits_0^1 \xi(\rho)\,\rho\,\rmn{d}\rho = \frac{1}{2}\,.
\label{eq:normalization}
\end{equation}
Let us consider the source passing a fold caustic
during the time\-span $2\,t_\star^\perp$ and let
$F_\rmn{f}^\star$ denote 
its flux at the beginning of a caustic entry or the end of a 
caustic exit at time $t_\rmn{f}^\star$.
In the vicinity of the caustic, the observed flux can be approximated
as \citep[e.g.][]{PLANET:sol,Do:Fold}
\begin{equation}
F(t) = F_\rmn{r}\,\left(\frac{\hat t}{t_\star^\perp}\right)^{1/2}\,G_\rmn{f}^\star\left(
\pm\,\frac{t-t_\rmn{f}^\star}{t_\star^\perp};\xi\right) + F_\rmn{f}^\star\,,
\end{equation}
where upper (lower) signs correspond to caustic entries (exits),
$F_\rmn{r}$ denotes a characteristic rise flux, $\hat t$ is an arbitrarily chosen
unit time, and the variation of images of the source that do not become critical as the
caustic is approached is neglected. The caustic profile function $G_\rmn{f}^\star(\eta;\xi)$ 
depends on the caustic passage phase $\eta = \pm\,(t-t_\rmn{f}^\star)/t_\star^\perp$
and is characteristic for the adopted stellar brightness
profile $\xi(\rho)$, where \citep{Hey:Fredholm}
\begin{equation}
G_\rmn{f}^\star(\eta;\xi) = 
\int\limits_{0}^{1} {\bmath{\mathcal T}} (\eta,\rho)\;\xi(\rho)\, \rmn{d}\rho\,,
\label{eq:response}
\end{equation}
and
\begin{equation}
{\bmath{\mathcal T}} (\eta,\rho) = 2\,\rho^{1/2}\,j\left(\frac{\eta-1}{\rho}\right)\,.
\label{eq:sensfunc}
\end{equation}
With $K$ denoting the complete elliptical integral of first kind, $j(z)$ reads
\citep{GG:detLD}
\begin{equation}
j(z) = \left\{\begin{array}{lcl}
0 & \mbox{for} & z \leq -1 \\
\frac{\sqrt{2}}{\upi}\,K\left(\sqrt{\frac{1+z}{2}}\right)
& \mbox{for} & -1 < z < 1 \\
\frac{2}{\upi\,\sqrt{1+z}}\,K\left(\sqrt{\frac{2}{1+z}}\right)
& \mbox{for} & z > 1
\end{array} \right.\,.
\end{equation}

\section{Power-law profiles}
A popular class of models for the stellar brightness profile $\xi(\rho)$
is a linear superposition of 
terms proportional to powers 
of $\cos \vartheta = \sqrt{1-\rho^2}$, where $\vartheta$ denotes
the angle between the normal to the stellar surface and the direction to the observer.
With the term proportional to $\cos^p \vartheta$ reading
\begin{equation}
\xi_{\{p\}}(\rho) = (1+p/2)\,(1-\rho^2)^{p/2}\,,
\end{equation}
the stellar brightness profile becomes
\begin{equation}
\xi(\rho; \Gamma_{\{p_1\}}\ldots \Gamma_{\{p_{k}\}}) = \
1+\sum_{i=1}^{k} \Gamma_{\{p_i\}}\,\left[\xi_{\{p_i\}}(\rho)-1\right]
\end{equation}
with $k$ coefficients $0 \leq \Gamma_{\{p_i\}}\leq 1$ corresponding
to the power $p_i > 0$ which moreover are
required to fulfill the
condition 
\begin{equation}
\sum_{i=1}^{k}  \Gamma_{\{p_{i}\}}  \leq 1\,.
\end{equation}

According to Eq.~(\ref{eq:response}), the caustic profile function $G_\rmn{f}^\star(\eta;\xi)$
enherits the linearity in $\Gamma_{\{p_i\}}$ from $\xi(\rho)$, making it a 
superposition
\begin{equation}
G^{\star}_{\rmn{f}}(\eta; \Gamma_{\{p_1\}}\ldots \Gamma_{\{p_{k}\}}) = \
1+\sum_{i=1}^{k} \Gamma_{\{p_i\}}\,\left[G^\star_{\rmn{f},\{p_i\}}(\eta)-1\right]
\end{equation}
of the caustic profiles functions
$G^{\star}_{\rmn{f},\{p_i\}}(\eta) \equiv G^{\star}_{\mathrm{f}}(\eta; \xi_{\{p_i\}})$
which correspond to the stellar brightness profiles $\xi_{\{p_i\}}(\rho)$, where
\begin{equation}
G^{\star}_{\rmn{f},\{p\}}(\eta) = \frac{1}{\sqrt{\upi}}\,
\frac{(1+\frac{p}{2})!}{(\frac{1+p}{2})!}\,
\int\limits_{\rm max(1-\eta,-1)}^{\rm max(1-\eta,1)}\frac{(1-x^2)^{\frac{1+p}{2}}}
{\sqrt{x+\eta-1}}\;
\rmn{d}x\,.\label{eq:Gfp}
\end{equation}

\section{Measurement of second coefficient}

\begin{figure}
\includegraphics[width=84mm]{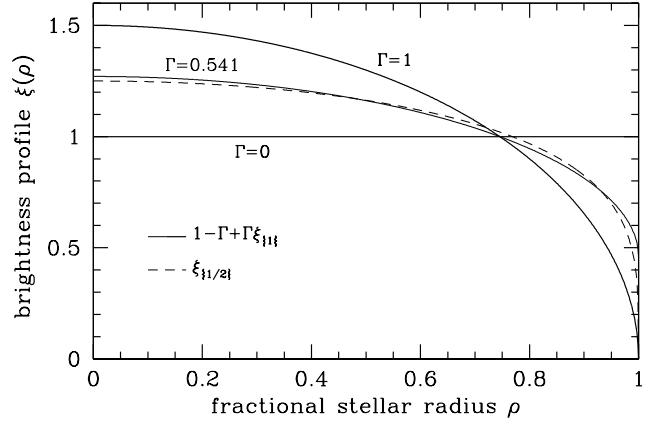}
\caption{Stellar brightness profiles $\xi(\rho)$ corresponding to different linear
superpositions of power-law models. The solid lines refer to
profiles $\xi(\rho; \Gamma) = 1-\Gamma 
+ \Gamma\, \xi_{\{1\}}(\rho)$ which are affine-linear
in $\cos \vartheta$, where $\Gamma = 0$ corresponds to
uniform brightness, $\Gamma = 1$ to maximal linear limb darkening, and
$\Gamma = 0.541$ to the best approximation in the sense of Eq.~(\ref{eq:norm}) to
maximal square-root limb darkening whose profile $\xi_{\{1/2\}}(\rho)$ is shown
as dashed line.}
\label{fig:intprofiles}
\end{figure} 

\begin{figure}
\includegraphics[width=84mm]{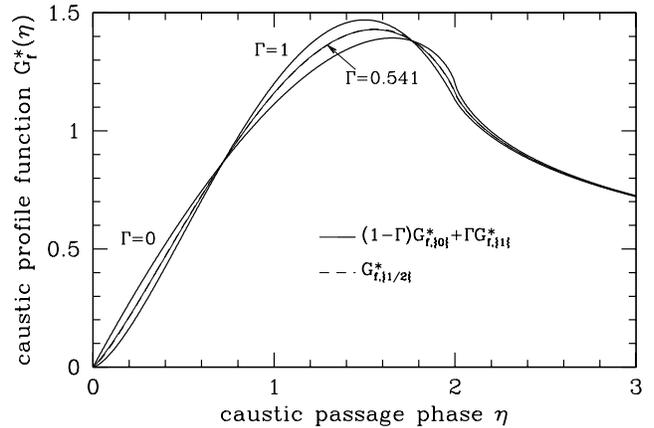}
\caption{Caustic profile functions $G^{\star}_{\rmn{f}}(\eta; \xi)$
that correspond to the stellar brightness profiles shown in Fig.~\ref{fig:intprofiles},
where $G^{\star}_{\rmn{f}}(\eta; \Gamma) = (1-\Gamma)\,G^{\star}_{\rmn{f}\,\{0\}}(\eta) 
+ \Gamma\,G^{\star}_{\rmn{f}\,\{1\}}(\eta)$ are plotted as solid lines for
$\Gamma = 0$ (uniform brightness), $\Gamma = 1$ (maximal linear limb darkening),
and $\Gamma = 0.541$ (best approximation to maximal square-root limb darkening), while the caustic profile function $G^{\star}_{\rmn{f},\{1/2\}}(\eta)$ 
for maximal square-root limb darkening
is shown as
dashed line.}
\label{fig:caustprofiles}
\end{figure} 

The stellar brightness profiles corresponding to uniform brightness ($p=0$), linear
limb darkening ($p=1$) and square-root limb darkening ($p=1/2$) 
are shown in Fig.~\ref{fig:intprofiles},
while the corresponding caustic profile functions are shown in Fig.~\ref{fig:caustprofiles}.
It can be seen that square-root limb darkening mediates between linear limb darkening and
uniform brightness. 

Let us compare maximal square-root limb darkening corresponding to 
the caustic profile function $G^{\star}_{\rmn{f},\{1/2\}}(\eta)$ with its
best approximation by a combination of uniform brightness and linear limb darkening
\begin{equation}
G^{\star}_{\rmn{f}}(\eta; \Gamma) = (1-\Gamma)\,G^{\star}_{\rmn{f}\,\{0\}}(\eta) 
+ \Gamma\,G^{\star}_{\rmn{f}\,\{1\}}(\eta)
\end{equation}
which depends on the choice of a norm on the function space. By requiring
\begin{equation}
\int\limits_0^\infty \left[G^{\star}_{\rmn{f}}(\eta; \Gamma)-
G^{\star}_{\rmn{f},\{1/2\}}(\eta)\right]^2 \rmn{d}\eta
\label{eq:norm}
\end{equation}
to adopt a miminum, the approximation is in maximal accordance with
the use of $\chi^2$ for sampled lightcurves
and yields $\Gamma = 0.541$ as optimal value. 
Figs.~\ref{fig:intprofiles} and~\ref{fig:caustprofiles} also show
the stellar brightness profile $\xi(\eta;\Gamma)$ and the caustic profile function
$G^{\star}_{\rmn{f}}(\eta; \Gamma)$, respectively, that corresponds to this value.
Since the caustic probes a combination of several fractional radii $\rho$ for any caustic
passage phase $\eta$ which is characterized
by ${\bmath{\mathcal T}} (\eta,\rho)$ according to Eq.~(\ref{eq:response}), significant differences between
the brightness profiles are washed out in the corresponding caustic profile functions making the
latter hardly distinguishable.

\begin{figure}
\includegraphics[width=84mm]{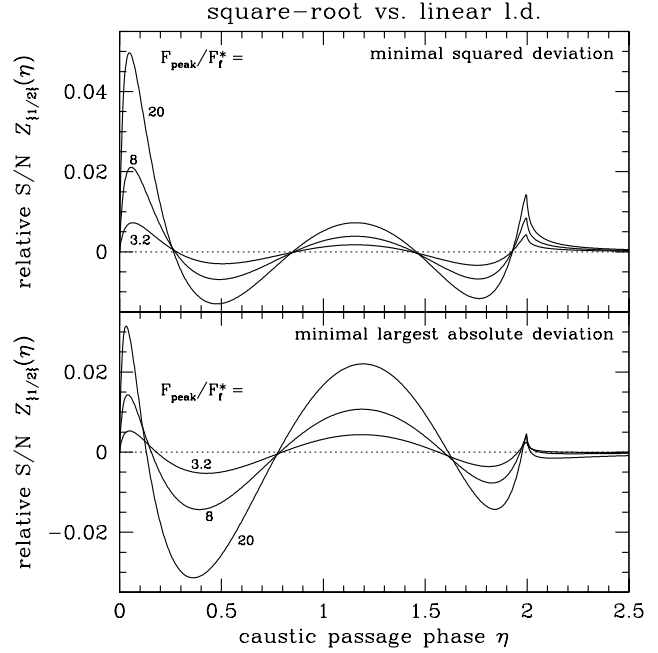}
\caption{Ratio $Z_{\{1/2\}}(\eta)$ between signal-to-noise $\Delta F(\eta)/\sigma$
for the detection of maximal square-root limb darkening
against the best-approximating affine-linear limb darkening and
signal-to-noise $F_\rmn{f}^\star/\sigma_\rmn{f}$ for the 
flux measurement outside the caustic
for selected caustic-peak-to-outside magnification ratios $F_\rmn{peak}/F_\rmn{f}^\star$.
In order to obtain the best approximation, $\Gamma$ has been chosen so that the integrated
square of 
$Z_{\{1/2\}}(\eta) = (\Delta F(\eta)/\sigma)/
(F_\rmn{f}^\star/\sigma_\rmn{f})$ is minimized for the curves shown in the upper
panel, yielding
$\Gamma =$0.542, 0.544, or 0.546 for  $F_\rmn{peak}/F_\rmn{f}^\star =$3.2, 8, or 20,
respectively,
whereas the absolute maximum of $Z_{\{1/2\}}(\eta)$ has been minimized for the curves shown in the
lower panel yielding $\Gamma =$0.567, 0.576, or 0.586 for the same values of 
$F_\rmn{peak}/F_\rmn{f}^\star$.}
\label{fig:signalsqrt}
\end{figure}

Let us assume that the uncertainties $\sigma$ of the flux measurements $F$
follow Poisson statistics, so that $\sigma = \sigma_\rmn{f}^\star\,\left(F/F_\rmn{f}^\star
\right)^{1/2}$, where $\sigma_\rmn{f}^\star$ denotes the uncertainty for the
flux $F_\rmn{f}^\star$ observed for the source being just outside the caustic.
The ratio $Z(\eta)$ between signal-to-noise for flux deviations $\Delta F$
due to the different profiles
and signal-to-noise for the flux $F_\rmn{f}^\star$ is then given by
\begin{eqnarray}
Z_{\{1/2\}}(\eta) & = & \frac{\Delta F(\eta)/\sigma}{F_\rmn{f}^\star/\sigma_\rmn{f}^\star} 
\nonumber \\  & = &
\alpha_{\{1/2\}}
\frac{G^{\star}_{\rmn{f}}(\eta; \Gamma)
-G^{\star}_{\rmn{f},\{1/2\}}(\eta)}{\sqrt{\alpha_{\{1/2\}}\,G^{\star}_{\rmn{f},\{1/2\}}(\eta) +
1}}\,,
\end{eqnarray}
where 
\begin{equation}
\alpha_{\{1/2\}} = \frac{(F_\rmn{peak}/F_\rmn{f}^\star) -1}{G_{\rmn{peak},\{1/2\}}}
\end{equation}
with $F_\rmn{peak}$ denoting the flux at the caustic peak corresponding to the maximum of
$G^{\star}_{\rmn{f},\{1/2\}}(\eta)$, which is $G_{\rmn{peak},\{1/2\}} \approx 1.431$.
For selected values of $F_\rmn{peak}/F_\rmn{f}^\star$, 
$Z_{\{1/2\}}(\eta)$ is displayed in Fig.~\ref{fig:signalsqrt}, where the linear limb-darkening
coefficient has been optimized so that either the integrated squared deviation or the
single largest absolute deviation is minimized. In general, the optimal choice of $\Gamma$
depends on the sampling, and with only parts of the caustic passage being sampled, the
signal may further decrease, also affected by adopting best-fit values for 
$t_\rmn{f}^\star$ and $t_\star^\perp$ that differ from the true values.
While a uniform sampling is better represented by the former choice for an optimal
linear limb-darkening coefficient, the latter choice better reflects the case where the
beginning of a caustic entry or the end of the caustic exit is preferentially sampled
being the region that provides the largest amount of information about the
square-root limb-darkening coefficient.

For all fold-caustic events observed so far, the caustic-peak-to-outside 
magnification ratio
is $F_\rmn{peak}/F_\rmn{f}^\star \la 8$, so that the measurement of 
$\Gamma_{\{1/2\}}$ with an absolute uncertainty of $0.1$ 
from a single point requires a relative flux uncertainty outside the
caustic $\sigma_\rmn{f}^\star/F_\rmn{f}^\star \la 0.2\,$\% for typical events, 
where this limit roughly increases proportional
to $F_\rmn{peak}/F_\rmn{f}^\star$.

\begin{figure}
\includegraphics[width=84mm]{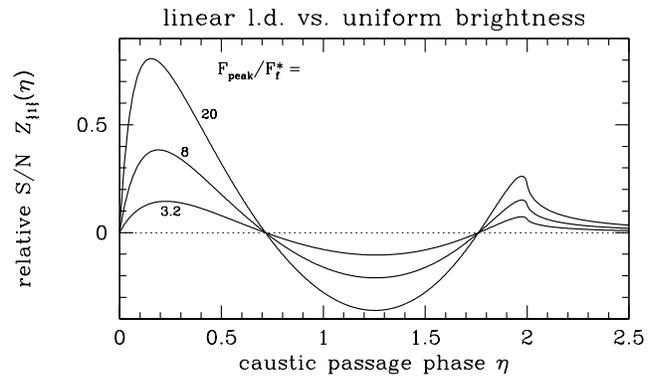}
\caption{Ratio $Z_{\{1\}}(\eta)$ between signal-to-noise $\Delta F(\eta)/\sigma$
for the detection of maximal linear limb darkening
against uniform brightness and signal-to-noise $F_\rmn{f}^\star/\sigma_\rmn{f}$ for the 
flux measurement outside the caustic for selected caustic-peak-to-outside 
magnification ratios
$F_\rmn{peak}/F_\rmn{f}^\star$.}
\label{fig:signallin}
\end{figure}

For comparison, Fig.~\ref{fig:signallin} shows the relative signal-to-noise ratio $Z(\eta)$
for the detection of linear limb darkening against uniform brightness, where
\begin{equation}
Z_{\{1\}}(\eta) = 
\alpha_{\{1\}}
\frac{G^{\star}_{\rmn{f},\{0\}}(\eta)
-G^{\star}_{\rmn{f},\{1\}}(\eta)}{\sqrt{\alpha_{\{1\}}\,G^{\star}_{\rmn{f},\{1\}}(\eta) +
1}}
\end{equation}
and
\begin{equation}
\alpha_{\{1\}} = \frac{(F_\rmn{peak}/F_\rmn{f}^\star) -1}
{G_{\rmn{peak},\{1\}}}\,,
\end{equation}
with $G_{\rmn{peak},\{1\}} \approx 1.470$ being the maximum of 
$G^{\star}_{\rmn{f},\{1\}}(\eta)$.

From a simulation of a caustic exit with $F_\rmn{peak}/F_\rmn{f}^\star = 12.5$ lasting
12~h and being sampled on average every 6~min with an accuracy 
$\sigma_\rmn{f}^\star/F_\rmn{f}^\star
= 1.5\,$\% in accordance with the capabilities of the PLANET campaign \citep{PLANET:EGS},
\citet{Do:FoldLD} found that a linear limb-darkening coefficient $\Gamma_{\{1\}} = 0.5$
can be measured with a relative uncertainty of $\sim\,2.5\,$\%. With the residual signal of an
additional square-root limb-darkening coefficient being $\sim\,$25 times smaller,
the absolute uncertainty in $\Gamma_{\{1/2\}}$ would be about 0.3 and not allow a meaningful
measurement. In order to achieve an uncertainty of 0.1 for $\Gamma_{\{1/2\}}$ in an event
with $F_\rmn{peak}/F_\rmn{f}^\star = 8$, one would therefore require a relative uncertainty
of the flux measurement outside the caustic of $\sigma_\rmn{f}^\star/F_\rmn{f}^\star \sim 0.3\,$\% and
less at the peak. With systematic effects dominating the photometric error bars rather than the
photon noise at this level, this appears not to be achievable. In addition
to large caustic-peak-to-outside magnification 
ratios $F_\rmn{peak}/F_\rmn{f}^\star$, large fluxes $F_\rmn{f}^\star$ at the
caustic outside also
enhance the prospects for measuring limb-darkening coefficients by allowing
better photometric accuracies $\sigma_\rmn{f}^\star/F_\rmn{f}^\star$.

\section{Mixed linear and square-root laws}

With square-root limb darkening being fairly approximated by a superposition of
linear limb-darkening and uniform brightness, 
photometric observations during the passage of 
a source star over a fold caustic appear
to be blind to the power index of an adopted limb-darkening law.

Assuming the presence of both linear and square-root limb darkening, it is
a linear
combination of the corresponding limb-darkening coefficients 
that is probed in first instance.

Let us consider that maximal square-root limb darkening $\xi_{\{1/2\}}(\rho)$
is best approximated by $\xi(\rho; \gamma)$, where $\gamma$ depends on the data.
The stellar brightness profile can then be written as
\begin{equation}
\xi(\rho) = 1 + \sum \left(\tilde \Gamma_{\pm}\,\tilde \xi_{\pm}(\rho) -1\right)\,,
\end{equation}
where
\begin{eqnarray}
\tilde \xi_{+}(\rho) & = & 
\frac{1}{1+\gamma}\left[\xi_{\{1/2\}}(\rho)+\gamma\,\xi_{\{1\}}(\rho)\right]\,, \nonumber \\
\tilde \xi_{-}(\rho) & = & 
\frac{1}{1-\gamma}\left[\xi_{\{1/2\}}(\rho)-\gamma\,\xi_{\{1\}}(\rho)\right]
\end{eqnarray}
fulfill the normalization of Eq.~(\ref{eq:normalization})
and analogous relations hold for $G^\star_\rmn{f}(\eta;\xi)$.
The mixed profiles $\tilde \xi_{+}(\rho)$                                       and $\tilde \xi_{-}(\rho)$ have been chosen so that $\tilde \xi_{+}(\rho)$
combines best-matched equivalent profiles, whereas $\tilde \xi_{-}(\rho)$
measures their difference.
The corresponding limb-darkening coefficients 
are then given by
\begin{eqnarray}
\tilde \Gamma_{+} & = & \frac{1+\gamma}{2\gamma}\,\left[\gamma\,\Gamma_{\{1/2\}} 
+ \Gamma_{\{1\}}\right] \,, \nonumber \\
\tilde \Gamma_{-} & = & \frac{1-\gamma}{2\gamma}\,\left[\gamma\,\Gamma_{\{1/2\}}  
- \Gamma_{\{1\}}\right] \,.
\end{eqnarray}
Therefore, the data will provide
an accurate measurement of $\tilde \Gamma_+$ similar to 
$\Gamma = \Gamma_{\{1\}}$ if only linear limb darkening is considered
(i.e.\ $\Gamma_{\{1/2\}} = 0$),
where one finds the correspondence 
$\tilde \Gamma_{+} = [(1+\gamma)/(2 \gamma)]\,\Gamma$ with $\gamma \sim$ 0.54 \ldots 0.60. 
With the influence of $\tilde \xi_{-}(\rho)$ on the lightcurve being
$\sim$\,25 times smaller than that of $\tilde \xi_{+}(\rho)$, the uncertainty
in $\tilde \Gamma_{-}$ corresponding to the second limb-darkening coefficient
will exceed that in $\tilde \Gamma_{+}$ by the same
factor, and models that   
coincide in $\tilde \Gamma_+$ are
roughly equivalent. 

With measurements of  $\tilde \Gamma_+$ and $\tilde \Gamma_-$, 
the coefficients corresponding
to the linear and square-root term follow as
\begin{eqnarray}
\Gamma_{\{1\}} & = & \frac{\gamma}{1+\gamma}\,\tilde \Gamma_{+} - 
\frac{\gamma}{1-\gamma}\,\tilde \Gamma_{-} \,, \nonumber \\
\Gamma_{\{1/2\}} & = & \frac{1}{1+\gamma}\,\tilde \Gamma_{+} +
\frac{1}{1-\gamma}\,\tilde \Gamma_{-} \,.
\end{eqnarray}

\section{Summary and further discussion}
If the stellar brightness profile $\xi(\rho)$
is modelled by a linear superposition of three base profile
functions, as for the discussed examples of constant, linear, and square-root terms
in $\cos \vartheta$, the normalization forces one of these profiles to mediate between the
other two, so that this base profile can be approximated by a superposition of the others.
With several fractional radii $\rho$ being probed by the fold 
caustic for any passage phase $\eta$
according to ${\bmath{\mathcal T}} (\eta,\rho)$, the residuals of this approximation are
smaller for the caustic profile function $G_\rmn{f}^\star(\eta; \xi)$ than for the
stellar brightness profile $\xi(\rho)$, which limits the power of 
photometric microlensing observations during fold-caustic
passages for revealing the stellar brightness profile $\xi(\rho)$.

Compared to the determination of a linear limb-darkening coefficient, 
the residual signals of 
an additional square-root limb-darkening coefficient 
are $\sim\,$25 times smaller, 
making its measurement with current microlensing follow-up
campaigns such as PLANET impossible, unless the caustic-peak-to-exit 
magnification ratio
becomes $F_\rmn{peak}/F_\rmn{f}^\star \ga 40$ for a (typical) photometric
accuracy  $\sigma_\rmn{f}^\star/F_\rmn{f}^\star
= 1.5\,$\% at the caustic exit 
where none of such events have been observed so far, or the source passes over a cusp
singularity rather than a fold.

If more than one limb-darkening base profile is assumed to contribute, 
only a single characteristic
coefficient can be accurately measured 
which corresponds to a specific superposition of the
corresponding
base profiles as found for worked examples \citep{PLANET:EB5,Abe}.

\section*{Acknowledgments}
This work has been made possible by postdoctoral support 
on the PPARC rolling grant
PPA/G/O/2001/00475. This letter provides the first part of 
the answer to a question asked by P.D.~Sackett several
years ago.

\bibliographystyle{mn2e}
\bibliography{sldcf}

\end{document}